\documentclass[a4paper,twocolumn,
english,aps,pre,floatfix,showpacs]{revtex4}
\usepackage[T1]{fontenc}
\usepackage[latin1]{inputenc}
\usepackage{amsmath}
\usepackage{babel}
\usepackage{graphics}
\usepackage{amssymb}


\begin{document}
\title
{Non-equilibrium processes: driven lattice gases, interface dynamics, and 
quenched disorder effects on density profiles and currents 
 }
\author {S.L.A. \surname{de Queiroz}}

\email{sldq@if.ufrj.br}

\affiliation{Instituto de F\'\i sica, Universidade Federal do
Rio de Janeiro, Caixa Postal 68528, 21941-972
Rio de Janeiro RJ, Brazil}

\author {R. B. \surname{Stinchcombe}}

\email{stinch@thphys.ox.ac.uk}

\affiliation{Rudolf Peierls Centre for Theoretical Physics, University of
Oxford, 1 Keble Road, Oxford OX1 3NP, United Kingdom}

\date{\today}

\begin{abstract}
Properties of the one-dimensional totally asymmetric
simple exclusion process (TASEP), and their connection with the dynamical
scaling of moving interfaces described by a Kardar-Parisi-Zhang (KPZ) equation
are investigated. 
With periodic boundary conditions, scaling of
interface widths (the latter defined via a discrete occupation-number-to-height 
mapping), gives the exponents 
$\alpha=0.500(5)$, $z=1.52(3)$, $\beta=0.33(1)$. 
With open boundaries, results are as follows:
(i) in the maximal-current phase, the exponents are the same 
as for the periodic case, and in agreement with recent Bethe ansatz results; 
(ii) in the low-density phase, curve collapse can be found to a rather 
good extent, with $\alpha=0.497(3)$, $z=1.20(5)$, $\beta=0.41(2)$, which is 
apparently at 
variance with the Bethe ansatz prediction $z=0$; (iii) on the coexistence line 
between low- and high- density phases,
$\alpha=0.99(1)$, $z=2.10(5)$, $\beta=0.47(2)$, in relatively good agreement with the
Bethe ansatz prediction $z=2$. From a mean-field continuum formulation,
a characteristic relaxation time, related to kinematic-wave propagation and having
an effective exponent $z^\prime=1$, is shown to be the limiting slow process
for the low density phase, which accounts for the
above-mentioned discrepancy with Bethe ansatz results. 
For TASEP with quenched bond disorder, interface width 
scaling gives $\alpha=1.05(5)$, $z=1.7(1)$, $\beta=0.62(7)$. 
From a direct analytic approach to steady-state
properties of  TASEP with quenched disorder, closed-form expressions for the 
piecewise shape of averaged density profiles are given, as well as rather 
restrictive bounds on currents. All these are substantiated in numerical simulations.

\end{abstract}
\maketitle
 
\section{Introduction} 
\label{intro}
In the absence, so far, of a general theory describing non-equilibrium processes
(even in steady state), it is worthwhile studying simple models whose 
qualitative features, it is hoped, will hold for a broad class of systems.    
In this paper we deal with properties of the one-dimensional totally asymmetric
simple exclusion process (TASEP)~\cite{derr98}, and their connection with the dynamical
scaling of moving interfaces described by a Kardar-Parisi-Zhang (KPZ)
equation~\cite{kpz,hhz95,bar95}. The TASEP is a biased diffusion process
for particles with hard-core repulsion (excluded volume)~\cite{derr98,sch00,derr93}.

Our main purpose is twofold: first, to probe the relationship between TASEP 
behavior and KPZ interface evolution under several distinct constraints, to be 
described below;
and, focusing especifically on systems with quenched disorder, to provide an account
of the effects of frozen randomness on the density profiles and currents in TASEP.

Upon implementing specific features on the particle system, such as various boundary 
conditions, assorted particle densities, currents, and/or injection/ejection rates, as 
well as quenched inhomogeneities, we measure the consequent changes to properties of the 
interface problem, taking the latter to be related to the former by the connection 
which we now sketch.
  
The $1+1$ dimensional TASEP is the fundamental discrete model for flow with 
exclusion. Here the particle number $n_\ell$
at lattice site $\ell$ can be $0$ or $1$, and the forward hopping of particles
is only to an empty adjacent site. Taking the stochastic attempt rate $p=1$
(see later) the current across the bond from $\ell$ to $\ell +1$ is thus 
$J_{\ell,\ell+1}= n_\ell (1-n_{\ell+1})$. This system maps exactly to an interface 
growth 
model~\cite{meakin86} in $D=1+1$, having integer values of height variables 
$h_{i(\ell)}$ on a new lattice such that $i(\ell)$ lies midway between sites $\ell$,
$\ell+1$ of the TASEP lattice. The $h_i$ are constrained by the relation
$h_{i(\ell+1)}-h_{i(\ell)}=1-2\,n_\ell$ ($=\pm 1$). When not concerned with
detailed associations of sites between the models we will omit the $(\ell)$ in
$i(\ell)$; after continuum limits, $(\ell)$ and $i(\ell)$ become the same.

There is clearly a precise one-to-one correspondence between these two discrete 
models and their stochasticity. Various associated continuum models are not
so clearly related, due to the possibility of different continuum 
limits. A naive continuum limit in the TASEP gives~\cite{ks91,kk08} the
noisy Burgers turbulence equation~\cite{ks91,fns77,vBks85} for one-dimensional particle 
flow:
this equation is the continuity equation resulting from a continuum
version of the TASEP bond current $J$, which takes the form
\begin{equation}
J= -\frac{1}{2}\frac{\partial
\rho}{\partial x} -\left(\rho -\frac{1}{2}\right)^2 +\zeta(x,t) +\frac{1}{4}\ ,
\label{eq:jdef}
\end{equation}
where $\rho=\langle n \rangle$ (i.e., the "mean field" version of $n$), and
$\zeta(x,t)$ is the (uncorrelated) noise here used to represent all the effects
of stochasticity: $\langle\zeta(x,t)\,\zeta(x^\prime,t^\prime)\rangle=\delta(x-x^\prime)\,
\delta(t-t^\prime)$.

Using the continuum "noiseless"  version of the height/occupation given above for the
discrete models, i.e., 
\begin{equation}
\frac{\partial h}{\partial x}=1-2\rho\ ,
\label{eq:corresp}
\end{equation}
it is seen that the Burgers equation is related by a spatial
derivative to the KPZ equation~\cite{kpz} for the evolution of the height $h(x,t)$
of an elastic interface above a fixed reference level:
\begin{equation}
\frac{\partial h}{\partial t}=\frac{\partial^2 h}{\partial x^2}+
\left(\frac{\partial h}{\partial x}\right)^2+\zeta(x,t)\ .
\label{eq:kpz}
\end{equation}

Depending on boundary conditions imposed at the extremities, one can have different 
regimes for the TASEP~\cite{derr98,derr93}, whose features will be recalled below
where pertinent. There should be corresponding regimes also for KPZ, as pointed out
previously~\cite{mukamel,kgmd93}.

In Section~\ref{sec:pbcss} we investigate the TASEP with periodic boundary conditions, 
and  the associated interface problem. By examination of interface width scaling, we 
estimate the respective critical exponents. We also study the behavior of asymptotic 
interface widths against particle density in the TASEP, as well as the main features of 
interface slope distributions and their connections to TASEP properties. In 
Sec.~\ref{sec:obcss}, we examine the interface width evolution corresponding to 
open-boundary TASEP systems in the following phases: (i) maximal-current, (ii) 
low-density, and (iii) on the coexistence line. In the latter case, we also calculate 
density profiles in the particle system. In Sec.~\ref{sec:qd} we turn to quenched bond 
disorder in the TASEP with PBC, providing a scaling analysis of the associated interface 
widths, as well as results for interface slope distributions. A direct analytic 
approach to steady-state properties of the TASEP with quenched disorder is given, 
focusing on averaged profiles densities and system-wide currents.
Finally, in Sec.~\ref{sec:conc}, concluding remarks are made.

\section
{Periodic boundary conditions}
\label{sec:pbcss}

We start by imposing periodic boundary conditions (PBC) for the TASEP at the ends of the 
chain, thus the total number of particles is fixed. Henceforth, the position-averaged 
particle density $\langle \rho\rangle$ will be denoted simply by $\rho$.
Several steady-state properties
are known exactly in this case~\cite{derr98}, as the configuration weights are 
factorizable at stationarity. To make contact with KPZ interface properties, we
consider the width $w (L,t)$ of an evolving interface of transverse size $L$ in $1+1$ 
dimensions in the discrete height model outlined in Section~\ref{intro}. 
The average interface slope is $1-2\rho$ [~see Eq.~(\ref{eq:corresp})~],  
and this average tilt must be taken into account. 
This is done by defining
\begin{equation}
\left[w(L,t)\right]^2
=L^{-1}\,\sum_{i=1}^{L}\left(h_i(t)-{h_1}(t)-(1-2\rho)\,(i-1)
\right)^2\ ,
\label{eq:width}
\end{equation}
where only fluctuations around the baseline trend are considered. 
Initially, $w$ grows
with time as $t^\beta$, until a limiting, $L$-dependent width $\sim L^\alpha$ is 
asymptotically reached. With $z=\alpha/\beta$, one expects from 
scaling~\cite{bar95,hhz95}:
\begin{equation}
w(L,t)=L^\alpha\,f\left(\frac{t}{L^{z}}\right)\ ,
\label{eq:wscale1}
\end{equation}
where
\begin{equation}
f(u)=\begin{cases}{u^\beta\quad u \ll 1}\cr{{\rm const.}\quad u \gg 1\ .}
\end{cases}
\label{eq:wscale2}
\end{equation}
For the $D=1+1$ KPZ model, one has the exact values~\cite{kpz}:
$\alpha=1/2$, $\beta=1/3$, $z=3/2$. 

\begin{figure}
{\centering \resizebox*{3.3in}{!}{\includegraphics*{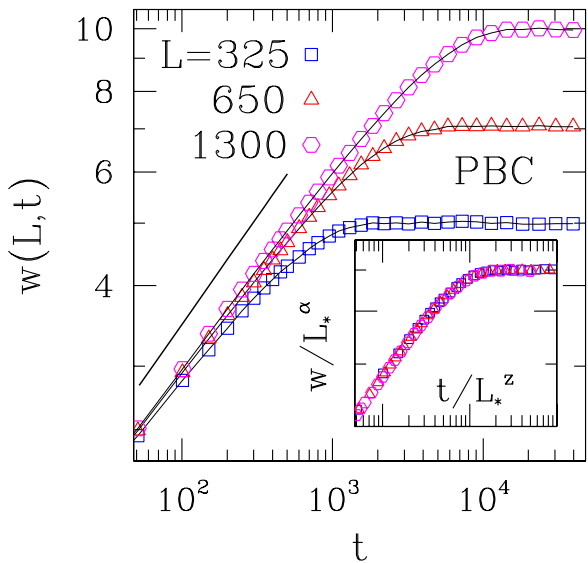}}}
\caption{(Color online) Double-logarithmic plot of interface width against 
time, corresponding to TASEP with periodic boundary conditions, $\rho=1/2$, lattice sizes 
$L$ as in key to symbols. Full line corresponds to $w \sim t^{1/3}$. Inset: scaling plot 
of data on main diagram, using $\alpha=1/2$, $z=3/2$; $L_\ast=L/325$.
} 
\label{fig:pbc}
\end{figure}

We have simulated the TASEP on lattices with
$L=325$, $650$, and $1300$ sites with PBC. For specified densities $\rho$, we
would start from a particle configuration as uniform as possible, in order to minimize 
the associated interface width. A time step is defined as a set of $L$
sequential update attempts, each of these according to the following rules: 
(1) select a site at random, and (2) if the chosen site is occupied and its neighbor to 
the right is empty, move the particle. Thus, the stochastic character resides exclusively 
in the site selection process. At the end of each time step we measured the width of
the corresponding interface configuration. We took typically $N_s=10^4$ independent 
runs, averaging the respective results for each $t$. The evolution of interface widths 
for $\rho=1/2$ and assorted lattice sizes is shown in 
Fig.~\ref{fig:pbc}; the goodness of their scaling with the exact KPZ indices is
typical of what is attained over the full density range. By examining the 
variation of data collapse quality against changes in the fitting exponents,
we estimate $\alpha=0.500(5)$, $z=1.52(3)$ for $\rho=1/2$. Direct measurement of the 
short-time
exponent $\beta$ is less accurate; for example, fitting the $L=1300$ data 
of  Fig.~\ref{fig:pbc} for $10^2 < t < 10^3$ to a single power law gives $\beta=0.31(1)$.

\begin{figure}
{\centering \resizebox*{3.3in}{!}{\includegraphics*{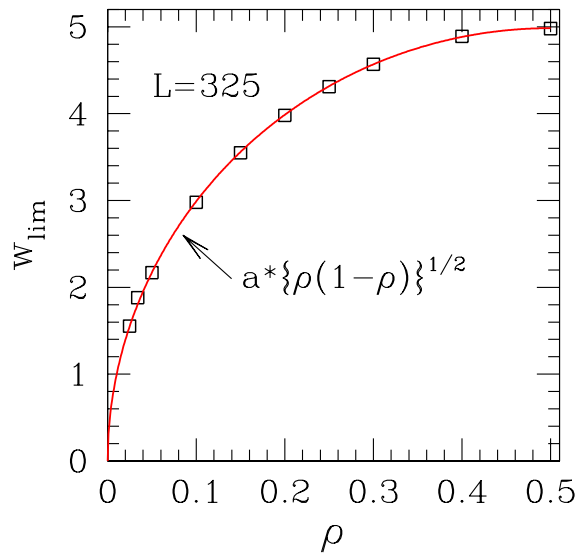}}}
\caption{(Color online) Limiting interface width $w_{\rm lim}$ against particle 
density $\rho$
for lattice with $L=325$, PBC. Squares are numerical results, from $N_s=10^4$ samples
each. Error bars are smaller than symbol sizes. The full line
is the analytic expression, Eq.~({\protect{\ref{eq:width3}}}), with
$a=9.97$ (see text).
} 
\label{fig:wvsrho}
\end{figure}

The limiting (asymptotic) interface width, $w_{\rm lim}$ obeys the $\rho \leftrightarrow 
1-\rho$ particle-hole symmetry of the TASEP, with a maximum at $\rho=1/2$. 
Fig.~\ref{fig:wvsrho} exhibits the behavior of $w_{\rm lim}$ for $\rho \leq 1/2$, and 
lattice size $L=325$. The leftmost data point corresponds to eight particles, i.e. $\rho 
=0.0246$. For even lower densities, discrete-lattice effects become more prominent, and 
the time needed to attain asymptotic behavior increases significantly. 

From Eq.~(\ref{eq:width}), using $h_i-h_{i-1}=1-2 n_{i-1}$, the squared width is
\begin{equation}
\left[w(L,t)\right]^2=\frac{4}{L}\,\sum_{i=1}^L\left\{(i-1)\rho-\sum_{m=1}^{i-1} n_m
\,\right\}^2\ .
\label{eq:width2}
\end{equation}
For the steady state with PBC, the configurational weights are 
factorizable~\cite{derr98}: each $n_m$ is 
independently distributed, taking values $(0,1)$ with probabilities $(\rho,1-\rho)$.
Consequently, in this case the average of $\left[(i-1)\rho-\sum_{m=1}^{i-1} n_m
\,\right]^2$ is $(i-1)(\rho-\rho^2)$, and the steady state rms width is
\begin{equation}
\left[\langle w(L)^2\rangle\right]^{1/2}=c\,(L-1)^{1/2}\,[\rho(1-\rho)]^{1/2}\ ,
\label{eq:width3}
\end{equation}
where $c$ is a numerical constant. This form, consistent 
with having the
KPZ scaling exponent $\alpha=1/2$, is compared with the simulation results
for averaged limiting widths in Fig.~\ref{fig:wvsrho}.

According to Eq.~(\ref{eq:corresp}), particle density fluctuations can be
investigated via the probability distribution functions (PDF) for slopes of the
associated interface problem. In doing so within the context of a 
discrete-lattice model, one must take recourse to a
coarse-grained description. We have experimented by taking interface segments
with a varying number $m$ of bonds, and calculating the average slopes between
the respective endpoints. For $m \gtrsim 10$ (the smallest practicable limit, such that 
the discreteness of allowed slope values still permits one to speak of a
relatively smooth PDF), the curves are very close to Gaussians, and their width (height 
at peak) varies with $m$ as $m^{-1/2}$ ($m^{1/2}$).

This can be understood by recalling the specific form of the (factorizable)
particle configuration weigths in this case~\cite{derr98}. Taking $\rho$
as the overall density, the probability 
of occurrence of a configuration with average slope $s=1-2x$ (i.e., average local
density $x \in [0,1]$), on a lattice section with $m$ sites is    
\begin{equation}
P(x) \sim C_m^{mx}\,\rho^{mx}\,(1-\rho)^{m(1-x)}\ .  
\label{eq:p(mx)}
\end{equation}
Standard treatment of Eq.~(\ref{eq:p(mx)}) shows that, close to the maximum
at $x=\rho$, the curve shape is indeed Gaussian, with a width proportional to
$m^{-1/2}$.

\begin{figure}
{\centering \resizebox*{3.3in}{!}{\includegraphics*{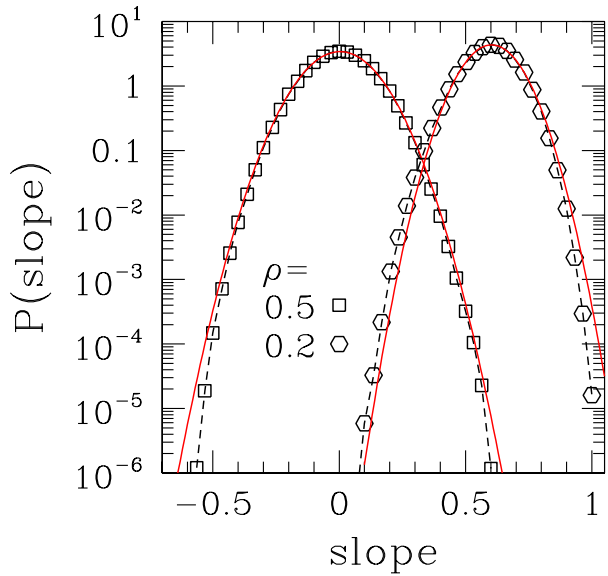}}}
\caption{(Color online) Slope PDFs in stationary state, 
lattice size $L=325$ with PBC, for $\rho=1/2$ and $0.2$. Full lines are 
Gaussian fits to data. 
} 
\label{fig:sldist}
\end{figure}

Having thus established
the nature of the $m$-dependence of slope PDFs, we have used $m=60$ in our
calculations, which gives a convenient, almost continuous, spectrum of allowed 
slopes. Results for the stationary state are depicted in Fig.~\ref{fig:sldist}.
One sees that the $\rho=0.5$ PDF is fitted by a Gaussian down to some four 
orders of magnitude below the peak, while for $\rho=0.2$, departures 
from a Gaussian profile are noticeable already at PDF values $\sim 10^{-2}$
times those at the maximum. 

\begin{figure}
{\centering \resizebox*{3.3in}{!}{\includegraphics*{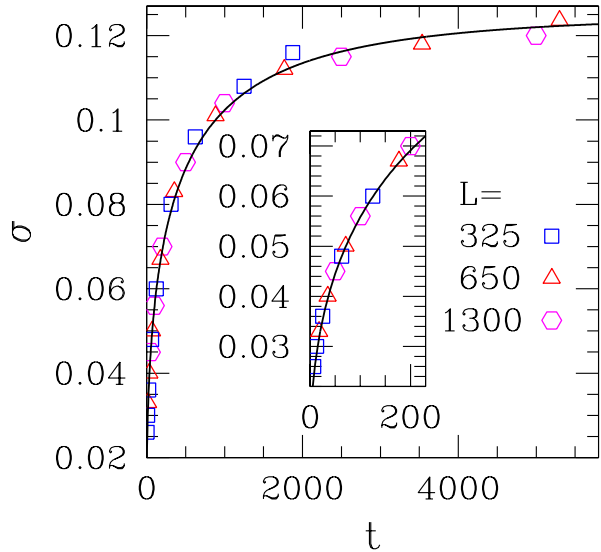}}}
\caption{(Color online) Width $\sigma$ of slope distributions in transient regime 
against time, for  $\rho=1/2$ and assorted lattice sizes $L$ with PBC. Initial 
particle configuration: alternate empty and filled sites, for minimum interface width.
Full line is stretched exponential fit to data (see text). Inset shows details of
main figure, close to the vertical axis. 
} 
\label{fig:slwvst}
\end{figure}

We have followed the evolution of interface slope PDFs during the transient regime
(starting from a particle distribution as uniform as possible), for $\rho=1/2$.
We ascertained that, already from early times, their shapes are very well 
approximated by Gaussians. Fig.~\ref{fig:slwvst} shows the (root-mean-square)
widths of Gaussian fits to the PDFs against time. 
It is noteworthy that, contrary to the behavior of interface widths shown in 
Fig.~\ref{fig:pbc}, here one does not find any significant dependence on lattice
size. It thus appears that, even during the transient, the range of density 
fluctuations for $\rho=1/2$ is shorter than the smallest lattice size considered 
here, $L=325$.
The solid curve in Fig.~\ref{fig:slwvst} is a stretched exponential,
$F(t)=a-b\,\exp\left\{-(t/t_0)^\delta\right\}$, for which the best-fitting 
parameter values are $t_0=380(20)$, $\delta=0.51(2)$. 

In summary, we have shown that our methods, namely inferring scaling properties of 
TASEP via those of the associated interface problem, do give rather accurate
results (where comparison is possible, i.e., for the scaling exponents $\alpha$,
$\beta$, and $z$) in the simplest case of a purely stochastic system with PBC. 
Our results so far are in agreement with the so-called KPZ conjecture~\cite{kk08},
sketched in Eqs.~(\ref{eq:jdef})-(\ref{eq:kpz}) above.

\section
{Open boundary conditions}
\label{sec:obcss}

\subsection{Numerical results}
\label{nr}

With open boundary conditions, the following additional
quantities are introduced: the injection rate $\alpha_I$ at the left end,
and the ejection rate $\beta_E$ at the right one (both defined as fractions
of the internal hopping rate). The number of particles is no longer constant, although
at stationarity it fluctuates around a well-defined average. 
Therefore, in order to consider the associated interface widths, one needs to subtract 
the instantaneous 
average slope $1-2\rho(t)$, in the manner of  Eq.~(\ref{eq:width}), at each time step.

Many stationary properties are known for this 
case~\cite{derr98,sch00,derr93,ds04,rbs01,be07}, including the phase diagram in  
$(\alpha_I, \beta_E)$ space. With open boundary conditions one must be aware that,
even at stationarity, ensemble-averaged quantities such as densities will 
locally deviate from their bulk values, within "healing" distances 
from the chain extremities which depend on the boundary (injection/ejection) rates.

Numerical density-matrix renormalization group (DMRG) techniques have been 
applied to estimate the dynamic exponent $z$ for several locations on the
phase diagram~\cite{nas02}. More recently~\cite{ess05}, a Bethe-ansatz solution 
has been provided, giving exact (analytic) predictions for the value of $z$  
everywhere on the phase diagram.

\begin{figure}
{\centering \resizebox*{3.3in}{!}{\includegraphics*{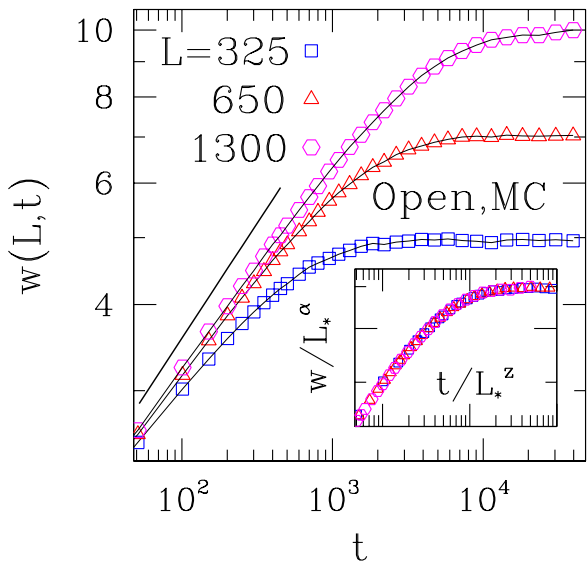}}}
\caption{(Color online) Double-logarithmic plot of interface width against 
time, corresponding to TASEP with open boundary conditions inside the maximal 
current (MC) phase, $\alpha_I=\beta_E=3/4$. Initial density $\rho=1/2$, lattice sizes 
$L$ as in key to symbols. Full line corresponds to $w \sim t^{1/3}$. Inset: scaling plot 
of data on main diagram, using $\alpha=1/2$, $z=3/2$; $L_\ast=L/325$.
} 
\label{fig:obcab75}
\end{figure}

We started by investigating the maximal-current (MC) phase ($J=1/4$) at
$\alpha_I > 1/2$, $\beta_E > 1/2$. The Bethe ansatz solution~\cite{ess05}
predicts the KPZ value $z=3/2$ there, concurrently with DMRG
results~\cite{nas02}. We took $\alpha_I =\beta_E =3/4$,
starting from an initial configuration with $\rho=1/2$ (in the present case, this is
the average final density as well), and alternate empty and 
filled sites, for minimum interface width. The evolution of interface widths against
time was very similar to the PBC case, as shown in Fig.~\ref{fig:obcab75}, and the KPZ
exponents $\alpha=1/2$, $\beta=1/3$, $z=3/2$ were extracted within error bars of the same 
order as those for PBC.
We also checked the multicritical point $\alpha_I =\beta_E =1/2$, and found the KPZ
exponents again there, with an accuracy similar to that obtained deep within the 
MC phase. The value $z=3/2$ has been found at this point by DMRG as 
well~\cite{nas02}.

Elsewhere on the phase diagram, a low-density phase exists at
$\alpha_I <1/2$, $\alpha_I <\beta_E$ (with $\rho=\alpha_I$), and a high-density phase at
$\beta_E<1/2$, $\beta_E <\alpha_I$ 
(with $\rho=1-\beta_E$)~\cite{derr98,sch00,derr93,ds04}. There is a critical coexistence 
line $\alpha_I=\beta_E < 1/2$, where a first-order transition occurs~\cite{derr98}. 
The low- and high-density
phases are further subdivided~\cite{ess05}. However, such subdivisions will not
concern us directly here, the relevant fact being that 
(outside the MC phase) the Bethe ansatz
solution predicts a non-vanishing gap as $L\to \infty$  (i.e. $z=0$) everywhere
(this is found by DMRG as well~\cite{nas02}), except 
on the coexistence line where diffusive behavior with $z=2$ is expected~\cite{ess05}.

\begin{figure}
{\centering \resizebox*{3.3in}{!}{\includegraphics*{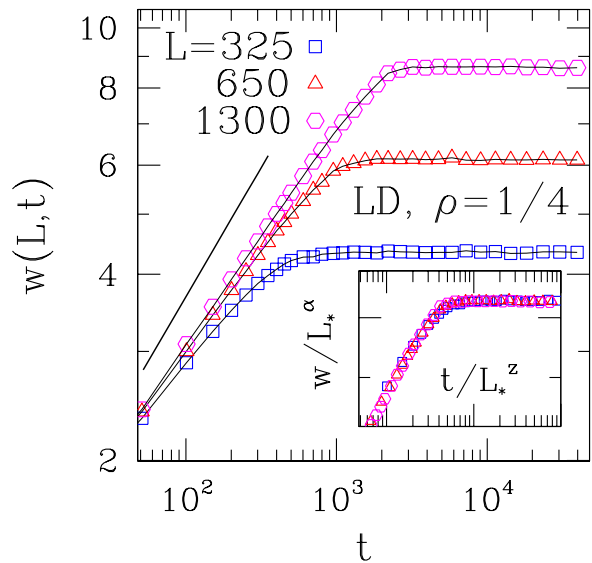}}}
\caption{(Color online) Double-logarithmic plot of interface width against 
time, corresponding to TASEP with open boundary conditions inside the low-density 
(LD) phase, $\alpha_I=1/4$, $\beta_E=1/2$. Initial density $\rho=1/4$, lattice sizes 
$L$ as in key to symbols. Full line corresponds to $w \sim t^{5/12}$. Inset: scaling 
plot of data on main diagram, using $\alpha=1/2$, $z=6/5$; $L_\ast=L/325$.
} 
\label{fig:obca25b5}
\end{figure}

We first examined a point away from the coexistence line, namely $\alpha_I=1/4$,
$\beta_E=1/2$ where the stationary density is thus $\rho=1/4$. Setting the initial 
density at the stationary value, the time evolution
of the associated interface widths is qualitatively very similar to the cases 
illustrated earlier, as shown in Fig.~\ref{fig:obca25b5}; however, scaling turns out 
to be very different.

Attempting curve collapse on the data of Fig.~\ref{fig:obca25b5} (see inset) 
gives the following estimates: $\alpha=0.497(3)$, $z=1.20(5)$.
This would imply $\beta=0.41(2)$ from scaling; 
a direct fit of $L=1300$ data for $10^2 < t < 10^3$ gives $\beta=0.35(1)$, which is a 
larger discrepancy compared with the scaling prediction than, e.g., for PBC or for the MC 
phase with open boundaries. 

These results for $\alpha$ and $z$ are inconsistent with the scaling relation 
$\alpha+z=2$ from galilean invariance~\cite{kpz}; however, it should be recalled that 
translational invariance, the key ingredient for galilean 
invariance to hold, is in general broken by the system's boundaries here present.
The fact that $\alpha+z=2$ is obeyed for open boundary conditions, in the MC  phase 
$\alpha_I$, $\beta_E \ge 1/2$, can be explained on the basis of simple kinematic-wave
theory~\cite{be07} for the TASEP. Indeed, in this phase the kinematic waves produced by 
both boundaries do not penetrate the system~\cite{be07}, and one does not see the 
formation of a shock (density wave) in the bulk which would otherwise disrupt the 
translational symmetry.

As mentioned above, both DMRG numerics~\cite{nas02} and the Bethe ansatz
solution~\cite{ess05} predict $z=0$, i.e., the correlation length is supposed to be 
finite here. 
We defer discussion, and a proposed solution, of the apparent 
contradiction between our own results and previous ones, 
to Subsection~\ref{mf} below, where a continuum mean-field
treatment of the approach to stationarity is developed. For the moment,
we note that an explanation can be provided for the limiting-width exponent $\alpha$,
by going back to 
Eq.~(\ref{eq:width2}) which was there applied to the factorizable-weight case of PBC.
As seen from the development immediately below Eq.~(\ref{eq:width2}), 
the only changes brought about by a finite correlation length (assumed to be $\ll L$) 
amount to an  $L$-independent correction to the term within round brackets, 
resulting from short-range density-density correlations. Thus, the $L^{1/2}$
dependence of $\left[\langle w(L)^2\rangle\right]^{1/2}$, given in
Eq.~(\ref{eq:width3}), remains valid here. This is corroborated by our
numerical estimate $\alpha=0.497(3)$, quoted above.

\begin{figure}
{\centering \resizebox*{3.3in}{!}{\includegraphics*{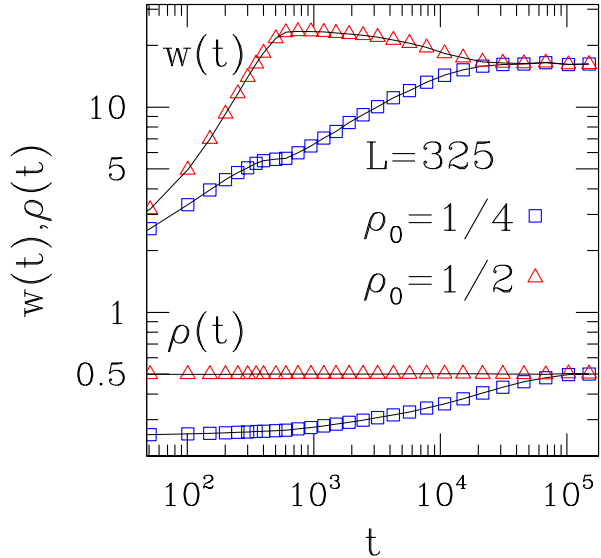}}}
\caption{(Color online) Double-logarithmic plot of interface width $w$ and overall 
particle density $\rho$ against  time, corresponding to TASEP with open boundary 
conditions, at $\alpha_I=\beta_E=1/4$ (on the coexistence line), lattice size 
$L=325$, and two distinct initial densities, $\rho_0=1/4$ and $1/2$.
} 
\label{fig:wrhoab25}
\end{figure}

Next, we looked at a point with $\alpha_I =\beta_E=1/4$, on the coexistence line.
In the stationary state, one expects a low-density phase with 
$\rho_-=\alpha_I$, and a  high-density one with $\rho_+=1-\beta_E$, to coexist.
We investigated the time evolution of both interface widths and
overall densities, with two different initial 
conditions, namely $\rho_0=1/4$, $1/2$ (because of particle-hole symmetry, starting
with $\rho=3/4$ gives the same interface width as for $\rho=1/4$, and a complementary
particle density). It can be seen in Fig.~\ref{fig:wrhoab25} that with both initial
conditions, the fixed point for the average density is $\rho=1/2$. While,
for $\rho_0=1/4$, the time evolution of the associated interface width still displays
the simple, monotonic, character found in all setups previously studied here (apart from 
a small bump at early times), such a feature is lost for $\rho_0=1/2$. Though the 
width  eventually  settles at a unique saturation value, the corresponding relaxation 
time is one order of magnitude longer than elsewhere on the phase diagram or for PBC (see 
the curves  for $L=325$ in the respective figures).

In order to unravel the corresponding spatial particle distributions, 
we also looked at the time evolution of slope PDFs at this point, for the same
initial densities. 

\begin{figure}
{\centering \resizebox*{3.3in}{!}{\includegraphics*{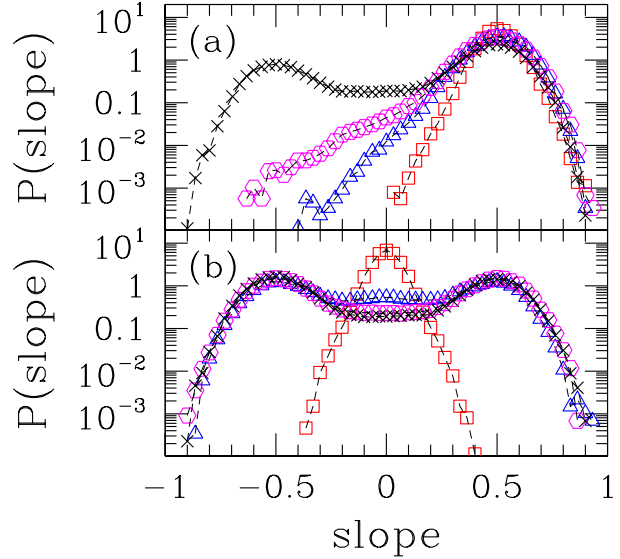}}}
\caption{(Color online) Slope PDFs at assorted times, for interfaces
corresponding to TASEP with open boundary  conditions, at $\alpha_I=\beta_E=1/4$ (on the 
coexistence line), lattice size 
$L=325$, and two distinct initial densities, namely (a) $\rho_0=1/4$, and  (b) 
$\rho_0=1/2$. Key to symbols: squares, $t=150$ (a), $t=50$ (b); triangles, $t=500$;
hexagons, $t=5000$; crosses, $t=20000$. 
} 
\label{fig:sldist2}
\end{figure}

Results are depicted in Fig.~\ref{fig:sldist2}, showing that
for both cases a double-peaked structure eventually evolves. The peak heights
indicate a large degree of spatial segregation between the $\rho_0=1/4$ and $3/4$ 
phases. For example, in case (b) [initial density $\rho_0=1/2$], at $t=20000$ the peaks
associated to the coexisting densities are $\sim 8$ times higher than the trough in 
the PDF at slope zero ($\rho=1/2$). 

\begin{figure}
{\centering \resizebox*{3.3in}{!}{\includegraphics*{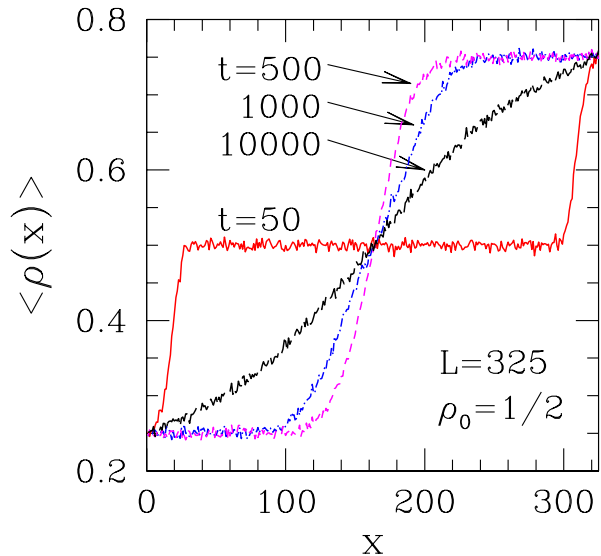}}}
\caption{(Color online) Ensemble-averaged particle density profiles at assorted times, 
for  TASEP with open boundary  conditions, at $\alpha_I=\beta_E=1/4$ (on the 
coexistence line), lattice size  $L=325$, and initial density $\rho_0=1/2$. Averages 
taken over $10^4$ independent samples (see text).
} 
\label{fig:prof}
\end{figure}

We pursued this point further, via direct examination of the evolution of averaged 
density  profiles in the particle system. Results for $L=325$, with initial density 
$\rho_0=1/2$  and initial distribution as uniform as possible (i.e., alternating empty and 
occupied sites), are shown in Fig.~\ref{fig:prof}. 

For $t \lesssim 500$, the low injection/ejection rates cause the 
the initial plateau of uniform density to be symmetrically eaten into,
as illustrated by the $t=50$ profile. After the two density waves meet,
a shock (kinematic wave) is formed, which for this case of $\alpha_I=\beta_E <1/2$,
is stationary on average~\cite{be07}, i.e., it jiggles around and is in effect
bounced off the system's boundaries. For the ensemble-averaged densities at fixed times
$\gtrsim 500$, the consequence of this is that the promediated profiles grow ever more
featureless (because the spread between locations of the shock, at the same time   
but for different noise realizations, increases with time owing to increasing 
sample-to-sample decorrelation). At $t=10000$, one sees
in Fig.~\ref{fig:prof} a nearly-constant slope, i.e., the local average density
increases roughly linearly with position along the system. 
Looking back at  Fig.~\ref{fig:wrhoab25}, however, it is apparent that a large degree of 
spatial segregation remains at stationarity between the $\rho=1/4$ and $3/4$ phases, 
within each realization. 

\begin{figure}
{\centering \resizebox*{3.3in}{!}{\includegraphics*{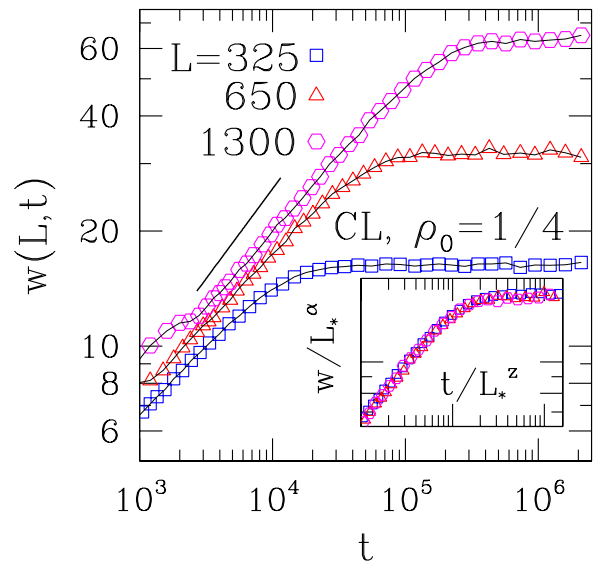}}}
\caption{(Color online) Double-logarithmic plot of interface width against 
time, corresponding to TASEP with open boundary conditions on the coexistence line, at 
$\alpha_I=\beta_E=1/4$. Initial density $\rho_0=1/4$, lattice sizes 
$L$ as in key to symbols. Full line corresponds to $w \sim t^{0.47}$. Inset: scaling plot 
of data on main diagram, using $\alpha=1$, $z=2.1$; $L_\ast=L/325$.
} 
\label{fig:obcab25}
\end{figure}

Finally, still on the coexistence line, we investigated the scaling properties of 
interface widths. For this,
we chose the initial condition $\rho_0=1/4$ which, as seen in 
Fig.~\ref{fig:wrhoab25}, results in a relatively smooth and monotonic evolution pattern.   
Our data (skipping the very early stage $t < 10^3$, for which the small bumps referred to 
above make their appearance) are displayed in Fig.~\ref{fig:obcab25}.

Attempting curve collapse on the data of Fig.~\ref{fig:obcab25} (see inset) 
gives the following estimates: $\alpha=0.99(1)$, $z=2.10(5)$.
This would imply $\beta=0.47(2)$ from scaling, as shown by the straight line on the
main plot; a direct fit of $L=1300$ data for $7\times 10^3 < t < 7\times 10^4$ gives 
$\beta=0.39(1)$, which again is a 
larger discrepancy compared with the scaling prediction than in cases previously
examined here.

The estimate of $z$ from curve collapse is roughly in line with the Bethe ansatz 
prediction~\cite{ess05} $z=2$, though it seems difficult to stretch the error
bars for our data to include this latter value. As regards $\alpha$, our estimate 
suggests that $\alpha=1$ is  possibly an exact result for this case.

An argument can be given as follows. Going back to the calculation outlined, for PBC, 
in Eq.~(\ref{eq:width2}), and recalling the spatial phase segregation shown
in the late-time data of Fig.~\ref{fig:sldist2}, one sees that the dominant feature
of the interface (particle system) is its division at a strongly fluctuating
interface into two main segments with symmetric 
slopes (low and high densities), of lengths $\propto L$. In order words, local
densities will be correlated along distances of order $L$. This is enough to
guarantee that the mean square interface width will depend on $L^2$.
This argument is supported by a calculation of
the size dependence of the stationary mean square width. On the
coexistence line the diffusing shock gives a linear average
stationary density profile (refer to the late-time curves in Fig.~\ref{fig:prof}), 
and this makes the large time limit of Eq.~(\ref{eq:width2}) proportional to 
$L^2\,(2\alpha_I - 1)^2$ , consistent with $\alpha = 1$.

\subsection{Continuum mean-field approach}
\label{mf}

We investigate how the effects of number conservation can influence the
approach to stationarity. 
We wish to find out whether, and under what conditions, transient 
(e.g., ballistic) phenomena can occur which would mask the underlying 
long-time relaxational dynamics.
Determination of the exponent $z$ via interface width scaling, as done in 
Secs.~\ref{sec:pbcss} and~\ref{nr}, relies heavily upon collapsing the 'shoulders'  which 
mark the final approach to stationarity, thus this method always picks out the longest 
characteristic time. Therefore one may ask, for the low-density phase
with open boundary conditions, whether the $L-$independent relaxation time implied
by the prediction $z=0$  is hidden underneath a longer process; 
and whether in this scenario the latter time, which is captured by interface 
width scaling, is associated with features not so far emphasised within 
Bethe ansatz investigations, as in the connection between ballistic motion 
and imaginary parts of energies. Such things might explain the seeming
mismatch between our results and those of Ref.~\onlinecite{ess05}. 

In the TASEP (and similar non-equilibrium
flow processes) the bulk hopping process conserves
total particle number - hence the continuity-equation nature of the
evolution equation, and the slow nature of long-wavelength
fluctuations of the density.  With open boundaries,
particles can appear and leave at the chain ends, thus causing the 
total number of particles inside to change. 

A group of particles crossing at a boundary immediately affects the
mean density and the local density near the boundary. Such a group
may be transferred into the interior as a kinematic wave/moving domain wall
provided the kinematic wave velocity $v$ is non-zero and of the
right sign. A time of order $L/v$ will be needed to affect the local density 
throughout a system of size $L$ (as for example is
typically required to achieve a steady state configuration).

This argument suggests that measurements of quantities (like ones
conserved in the interior) whose changes are boundary-induced will
show characteristic times of order $L^{\,z^\prime}$, with $z^\prime$ at least 
unity (still slower limiting processes may also be involved, owing to
intrinsic features of the dynamics).

In order to provide a quantitative counterpart to these ideas, we have
used a continuum mean-field approach (see, e.g., Ref.~\onlinecite{rbs01})
in which the system is described by the noiseless Burgers/KPZ equations, 
linearized by the Cole-Hopf transformation~\cite{hopf50,cole51}. 
This picks up the kinematic wave effects in the evolution of
the local density $\rho_\ell$ and its mean $\rho$. It captures the ballistic
transport between boundary and interior, described qualitatively
above, which gives $z^\prime=1$.
In this description the reduced density $\sigma_\ell(t)=\rho_\ell(t) -1/2$
evolves like:
\begin{eqnarray}
\sigma_\ell(t)=
\frac{\partial}{\partial\ell}\,\ln \{\cosh K(\ell -\ell_0)\,
e^{-K^2t} +
\nonumber \\
+{\sum_k}^\prime ( A_k\,e^{k\ell}+ B_k\,e^{-k\ell})\,e^{-k^2t}\}
\ ,
\label{eq:sig_ell}
\end{eqnarray}
towards a steady state:
\begin{equation}
\sigma_\ell = K\,\tanh K(\ell-\ell_0)\ ,
\label{eq:sig_st}
\end{equation}
where $K$ and $\ell_0$ are decided by boundary rates 
($\alpha_I,\beta_E$), $K$ being real, except in the MC phase where it is
imaginary. The sum ${\sum}^\prime$ takes care of the difference between
initial and steady states, and (since this difference is decaying)
it involves complex $k$ with ${\rm Re}\,(k^2) > K^2$; this $k$ has to be
consistent with boundary conditions. $A_k$, $B_k$ are therefore
determined by initial conditions. 
The difference typically becomes a
kinematic wave  or domain wall~\cite{sch00,be07,lw55,djls03,ksks98,ps99,hkps01,sa02}.
If it is kink--like, its velocity $v$ depends as follows on
coarse-grained densities $\rho_<$, $\rho_>$ and currents $J_<$, $J_>$
on either side of the kink: $v=(J_< - J_>)/(\rho_< - \rho_>)$.
These quantities are set by the steady-state and initial (uniform)
density profiles; for smooth small--amplitude kinematic waves
$v=dJ/d\rho = 1 - 2 \rho$, and then the local coarse-grained density
$\rho$ is at late times set by the steady state.

For our investigations at $(\alpha_I, \beta_E) = (1/4, 1/2)$ the
initial profile $\rho_\ell = 1/4$ has to go into the steady-state profile,
which corresponds to $K=1/4$, $\ell_0 = L$, and
differs from the initial one only in having an upturn (to $\rho_L = 3/8$) 
at the right  boundary.

The kinematic wave velocity for $\rho = 1/4$ is $1/2$, i.e., positive, so
there is ballistic transfer, actually of vacancies at speed $1/2$ from the right
boundary, which after time $t \simeq L/v = 2L$ produces the steady-state
profile with its upturn at the right boundary. This is corroborated by our numerical
results depicted in Fig.~\ref{fig:obca25b5}. For all cases $L=325$, $650$, and 
$1300$, the interface width at $t=2L$ has reached  more than $98\%$ of its 
asymptotic value. Thus, while attempts at producing overall curve collapse indicate 
$z^\prime \simeq 1.20$ as quoted above, by focusing exclusively on the scaling of the 
'shoulders' one gets a result much closer to the mean-field value.

We also applied this picture to other cases investigated in Subsection~\ref{nr},
to check for consistency. Results are as follows.

For the MC phase at $(\alpha_I, \beta_E) = (3/4, 3/4)$,
initially $\rho_\ell = 1/2$, so $v=0$ and the steady state $\rho_\ell$
differs from the average by an upturn to $2/3$ at the left boundary,
and a downturn to $1/3$ at the right one. So ballistic effects are not very 
important; furthermore the limiting process is the slower "KPZ diffusion" 
having $z=3/2$.

On the coexistence line at $(\alpha_I, \beta_E) = (1/4, 1/4)$,
two initial densities were used: (a) $\rho_\ell = 1/2$, (b) 
$\rho_\ell = 1/4$.
For (a), the essential aspects are all as given in Fig.~\ref{fig:prof},
i.e., a ballistic early evolution from the domain walls coming in
from either end, followed by the limiting slower diffusive process ($z=2$).
These together give rise to the non-monotonic form in Fig.~\ref{fig:wrhoab25}.
For (b),  the ballistic process is effective for a long time (of order $L$) 
during which the average $\rho$ builds up, but again
the limiting process is diffusion ($z=2$).

So, the mean-field continuum picture is consistent with all our numerical results
for open boundary conditions, especially the one where we differ from 
Ref.~\onlinecite{ess05}. In this latter case, the characteristic time arising from
propagation of the kinematic wave is longer than the intrinsic ($L-$independent) one,
thus resulting in the effective exponent $z^\prime=1$ (for all other cases,
an exponent $z>1$ associated to intrinsic dynamics dominates anyway). 
Note that, 
for this scenario to work, one needs the kinematic wave to have non-zero
effective velocity $v$, and (for a characteristic relaxation time, proportional to $L$,
to show up) one also needs the kinematic wave to have to traverse a length of
order $L$ -- whether that is the case depends on the maximum 
distance from the nearest effective boundary to the place(s) where the profile has to be
adjusted.
 
In summary, the difference between our results and those of Ref.~\onlinecite{ess05}
is real, and interpretable as above. The effects and interpretation may be important for 
other quantities conserved in the bulk which build up only by propagation from the
boundaries.

\section
{quenched disorder}
\label{sec:qd}
Quenched disorder in the TASEP has been studied by many 
authors~\cite{tb98,bbzm99,k00,ssl04,ed04,hs04}. Here, we investigate bond 
disorder~\cite{k00}, i.e., while all particles are identical, the site-to-site hopping
rates are randomly distributed. 

Hereafter, we restrict ourselves to binary distributions $P(p)$ for
the internal nearest-neighbor hopping rate $p$: 
\begin{equation}
P(p)=\varphi\,\delta(p-p_s)+(1-\varphi)\,\delta(p-p_w)\ ,  
\label{eq:distp}
\end{equation}
where $p_s >1/2 $, $p_w <1/2 $ are associated, respectively, to "strong"  and "weak" 
bonds.

One must recall that
the bond disorder introduced above gives rise to correlated, or "columnar"
disorder in the associated interface problem~\cite{k00,ahhk94}. Indeed, the (fixed) value
of the hopping rate at a given position $x$ along the particle-model axis will
determine, once and for all, the probability of the height $h(x)$ being updated.
This is in contrast with the usual picture of quenched disorder in the KPZ
model~\cite{l93,csahok93,opz95,amaral95,sak02,katsu04}, where it is assumed
that the intensity of disorder is a random function of the instantaneous
two-dimensional position $\left(x,h(x)\right)$ of the interface element at $x$.

\begin{figure}
{\centering \resizebox*{3.3in}{!}{\includegraphics*{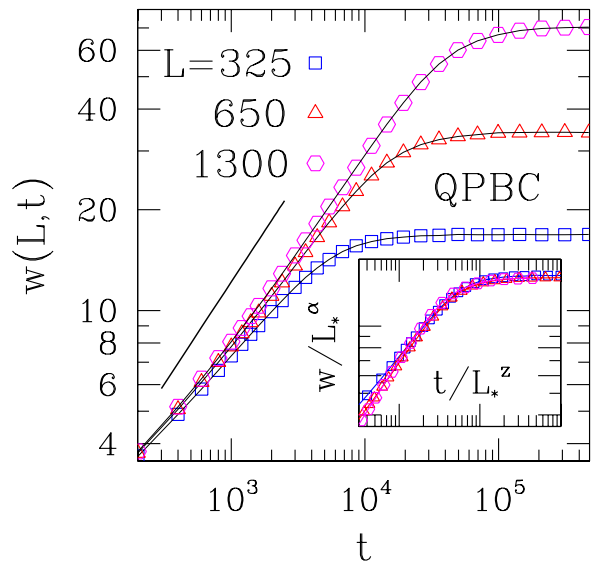}}}
\caption{(Color online) Double-logarithmic plot of interface width against 
time, corresponding to TASEP with periodic boundary conditions and binary quenched 
disorder, with $\varphi=1/2$, $p_s=0.8$, $p_w=0.2$ [see Eq.~(\protect{\ref{eq:distp}})]. 
$\rho=1/2$, lattice sizes  $L$ as in key to symbols. Full line corresponds to $w \sim 
t^{0.61}$. Inset: scaling plot 
of data on main diagram, using $\alpha=1.05$, $z=1.7$ (see text); $L_\ast=L/325$.
Each point is an average over $10\,000$ independent realizations of quenched randomness.
} 
\label{fig:qpbc}
\end{figure}

In Figure~\ref{fig:qpbc} we show interface width data for PBC, $\rho=1/2$, and 
$\varphi=1/2$, $p_s=0.8$, $p_w=0.2$ in Eq.~(\ref{eq:distp}). Note that the relaxation 
times are one order of magnitude longer than is usual for pure systems (in the
latter case, one must make exception for the coexistence line with open boundary 
conditions). Though the overall picture of a scaling regime still holds, with
interface widths evolving in a simple, monotonic way,
in general the quality of data collapse is lower than for pure
systems. From our best fit, we estimate $\alpha=1.05(5)$, $z=1.7(1)$, from which
scaling gives $\beta=0.62(7)$. A direct fit of $2\times 10^3 \leq x \leq 2\times 10^4$
data for $L=1300$ gives $\beta=0.56(1)$, only just within the error bars predicted by 
scaling. 

\begin{figure}
{\centering \resizebox*{3.3in}{!}{\includegraphics*{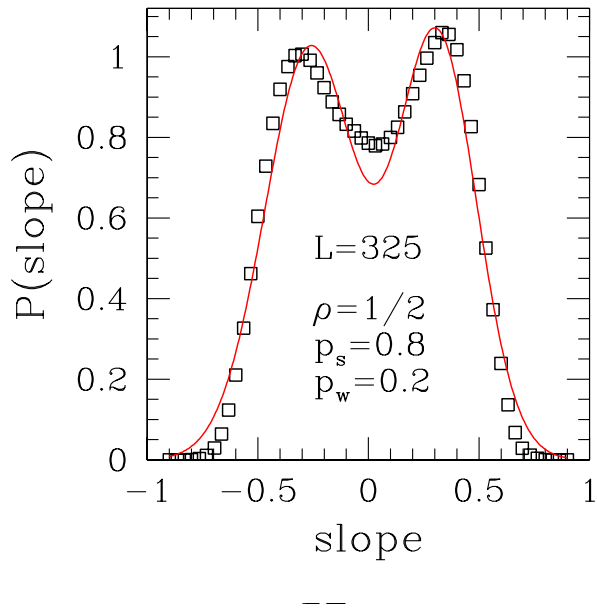}}}
\caption{(Color online)
Slope PDF in stationary state, corresponding to TASEP with quenched disorder,
 $\varphi=1/2$, $p_s=0.8$, $p_w=0.2$ [see Eq.~(\protect{\ref{eq:distp}})]. 
Lattice size $L=325$ with PBC, $\rho=1/2$. Full line is a double 
Gaussian fit to data (see text). 
$100$ independent realizations of quenched randomness were used, for each of
which samples were 
collected from $3\times 10^4$ consecutive interface configurations. 
} 
\label{fig:sldistq}
\end{figure}

We have examined the stationary-state slope distributions for the interface problem.
Numerical results are displayed in Fig.~\ref{fig:sldistq}, together with their fit by
a double-Gaussian form, $\Phi(x)=a\,G_1(x)+(1-a)\,G_2(x)$, where $G_i$ is a Gaussian
centered at $x_i$ with variance $\sigma_i^2$. The best-fitting parameters give a 
roughly symmetric curve, with $a=0.53(2)$,
$x_1=-0.26(1)$, $x_2=0.31(1)$, $\sigma_1=0.21$, $\sigma_2=0.18$. Such a double-peaked
structure indicates phase separation, as seen earlier for pure systems on the 
coexistence line (though here this is quantitatively milder, as the peak-to-trough 
ratio is $\sim 1.3$, to be compared to $\sim 8$ in the previous case). Phase separation 
is known to be a feature of the quenched-disordered TASEP~\cite{tb98}.

We now outline a theoretical framework for the description of the quenched-disordered 
problem, in the spirit of earlier work by Tripathy and Barma~\cite{tb98}. 
The analytic understanding of slope distributions and, particulary, of currents
and profiles, rests on the division of the system, for a given disorder
configuration, into an alternating succession of weak-bond and strong-bond segments
(${\cal S}^w$ and ${\cal S}^s$) each containing only bonds of one strength. The
longest weak bond segments determine the current through the system. This is easily
seen in a mean field account of the steady state of the particle system (with PBC).
Here the constant current $J$ yields the relation (profile map) 
\begin{equation}
\rho_{\ell+1}= 1 -\frac{J}{p_\ell\,\rho_\ell}\ ,
\label{eq:pmap}
\end{equation}
where $p_\ell$ is the hopping rate from site $\ell$ to $\ell+1$, and $\rho_\ell$
is the mean occupation of site $\ell$.

Throughout a weak bond segment ${\cal S}^w_n$
of length $n$ the map involves the constant (reduced) current $J/p_\ell=J/p_w \equiv 
J_w$. The corresponding strong bond
variable is  $J/p_s \equiv J_s$; $J_s < J_w$ since $p_s > p_w$. So, within any  ${\cal 
S}^w_n$  (or ${\cal S}^s_n$) the profile map is that of an effective pure system,
which is well known to give density profiles of kink shape, corresponding to low current
or high current: $\rho_\ell -\frac{1}{2}=k\,\tanh k(\ell-\ell_0)$ (monotonic increasing)
or $\rho_\ell -\frac{1}{2}=-K\tan K(\ell-\ell_0)$ (monotonic decreasing), depending
on whether the reduced current $J/p$ is less than or greater than $1/4$. $k$ and $K$
are related to the reduced current by $k=\sqrt{(1/4)-(J/p)}$;  
$K=\sqrt{(J/p)-(1/4)}$~\cite{hs04}. 

In the high current case $K$ has to be small [$\,\lesssim 
{\cal O}(1/L)$ in a segment of size $L\,$], to prevent the tangent from
diverging and taking $\rho_\ell$
outside of the permitted physical range $[0,1]$. In the binary random system, it is not
possible to have both weak and strong bond segments 
(having respectively $K=K_w$, $K_s$) in the high current 'state'
since that would lead to monotonically 
decreasing $\rho$ for all segments. That would violate the periodic boundary condition
requirement. It cannot even apply with open boundary conditions in a large system, 
because even with $K_s$ kept small by having $J_s$ close to $1/4$, the larger $J_w$
would cause a large $K_w$, resulting in non-physical values of $\rho_\ell$.
Arguments of this sort show that the strong bond segments ${\cal S}^s$ are all
in the low current phase, i.e., $J_s <1/4$ with density profiles in each
${\cal S}^s$ increasing monotonically. Thus, for PBC the profiles in 
each ${\cal S}^w$ have to decrease. That can come about from having $J_w >1/4$,
which leads to $K_w=\sqrt{J_w-(1/4)}$. If the longest weak bond segment has length $n_0$,
to prevent unphysical $\rho_\ell$'s resulting from a divergence of $\tan K_w 
(\ell-\ell_0)$ somewhere within that segment we must have $K_w \leq \pi/n_0$.
This gives:
\begin{equation}
\frac{1}{4} \leq \frac{J}{p_w} \lesssim \frac{1}{4} + \left(\frac{\pi}{n_0}\right)^2\ 
.
\label{eq:jwbound}
\end{equation}
It is straightforward to show that the characteristic length of weak bond segment is
$n \sim \ln\left(1/(1-\varphi)\right)$, and the largest weak segment length is
$n_0 \sim\ln L/ \ln\left(1/(1-\varphi)\right)$, so the above condition on $J_w$
is very restrictive, making the current typically $J \sim(p_w/4)+{\cal O}
\left((1/\ln L)^2\right)$.

This is in agreement with simulation results $J \simeq 0.055$ and $0.029$, 
respectively for $p_w=0.2$ and $p_w=0.1$ (both with $\rho=1/2$ on a lattice with 
$L=325$). 

\begin{figure}
{\centering \resizebox*{3.3in}{!}{\includegraphics*{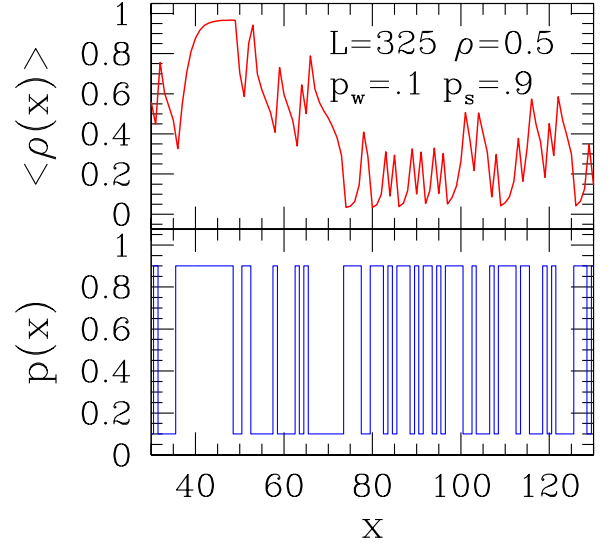}}}
\caption{(Color online) Upper part: time-averaged particle density against position
on section of $L=325$ system with PBC, corresponding to TASEP with quenched disorder, 
for a fixed bond configuration (seen in lower part of figure). Overall density 
$\rho=1/2$,  $\varphi=1/2$, $p_s=0.9$, $p_w=0.1$ [see Eq.~(\protect{\ref{eq:distp}})]. 
Average is over $4.65\times 10^5$ successive time steps, in stationary state.
Note kink-shaped profile for $35 \lesssim x \lesssim 45$, in direct correspondence
with the longest concentration of strong bonds; also, the high-current profile
for  $65 \lesssim x \lesssim 75$, coinciding with the longest concentration of
weak bonds.  
} 
\label{fig:pro325rho5q}
\end{figure}

A decreasing profile in each ${\cal S}^w$ is also possible with $J_w$ slightly less
than $1/4$ since, in addition to the kink-shaped low current profile, a profile with
$\rho_\ell$ decreasing between $1$ and $(1/2)+k$ can result from the low current map.
Again the $K_w$ that allows that is limited, leading for this case to 
$J \sim(p_w/4)-{\cal O}\left((1/\ln L)^2\right)$. This was seen, for example, in 
simulations for 
$\rho=0.8$, where $p_w=0.1$ gives $J \simeq 0.024$. This same result can also arise
from another low current profile (having $\rho_\ell$ decreasing between $(1/2)-k$ and 
zero). All these weak segment profiles are seen in simulation results, together with 
the characteristic kink-shaped profiles of the strong-bond segments (e.g., for
$p_w=0.1$, $\rho=0.5$ see Fig.~\ref{fig:pro325rho5q}, having $J =0.0297(3)$,
i.e., $J/p_w$ just greater than $1/4$).

Local density or slope distributions $P(\sigma)$, with $\sigma \equiv \partial 
h/\partial x$, are available from the profiles $\rho_\ell$, just as for the pure case,
by using the known density-slope relation, Eq.~(\ref{eq:corresp}) in:
\begin{equation}
P(\sigma)=\sum_{\rm segments} \int d\sigma_\ell\,\frac{\delta (\sigma -\sigma_\ell)}
{|\,d\sigma_\ell/d\ell\,|}=\sum_{\rm segments} 
\frac{1}{|\,d\sigma_\ell/d\ell\,|}_{\sigma_\ell=\sigma}\ .
\label{eq:p(sigma)}
\end{equation}
Contributions from high current segments give (using $\sigma_\ell=-K \tan K\ell\,$):
\begin{equation}
C \equiv \frac{1}{|\,d\sigma_\ell/d\ell\,|}_{\sigma_\ell=\sigma} = 2\left[ 
\sigma^2 +4K^2\right]^{-1}\ ,
\label{eq:p(K)}
\end{equation}	
while low current segments have contributions
\begin{equation}
C=\begin{cases}{
2\left[4k^2-\sigma^2\right]^{-1}\quad {\rm for}\ |\,\sigma|<2k}\cr
{2\left[\,\sigma^2-4k^2\right]^{-1}\quad {\rm for}\ |\,\sigma|>2k\ .}
\end{cases}
\label{eq:p(k)}
\end{equation}
In Eqs.~(\ref{eq:p(K)}),~(\ref{eq:p(k)}), these contributions are superimposed with
$\varphi$--dependent weights related to their frequency of occurrence. Stochastic 
effects,
missed from this account, cause fluctuations in the spatial pinning of the profiles
(as for the pure case), leading mainly to a smoothing of the divergences near 
$\sigma=\pm 2k$
in the low current segment contributions. Quantitative comparison of these 
predictions, e.g., to the data displayed in Fig.~\ref{fig:sldistq}, is not entirely 
appropriate,
on account of the coarse-graining introduced by taking slope samples along segments
of size $m$ bonds (typically $m=60$, as mentioned above). Nevertheless, one can see
that some general aspects, such as the appearance of peaks roughly equidistant 
from $\sigma=0$, are effectively mirrored in the numerical results.    

\section{Discussion and Conclusions} 
\label{sec:conc}
We start our discussion by recalling that the thermodynamic limits of
interface models, such as KPZ, and of exclusion processes, differ in
a subtle way. Namely, as can be seen from Eqs.~(\ref{eq:wscale1})
and~(\ref{eq:wscale2}), if one takes the $L \to \infty$ limit before
allowing $t \to \infty$, one is left with a perpetual coarsening transient in
which the interface gets ever rougher. On the other hand, the TASEP 
displays a well-defined stationary state even on an infinite system.
Therefore, the correspondence of the limiting-width regime of the
former type of problem to the stationary state of the latter can
be only take effect within a finite-size scaling context.

Once one is mindful of this distinction, however, the similarities and differences 
between two finite systems linked by the correspondence recalled in 
Sec.~\ref{intro} are expected to be {\it bona fide} features, which reflect
objective connections between the physics of non-equilibrium flow
processes, and that of moving interfaces in random media.

In Sec.~\ref{sec:pbcss} we first confirmed that the known KPZ exponents can be
numerically extracted, with good accuracy, from the scaling of
interface widths derived from the underlying TASEP with PBC. 
For a given range of system 
sizes, direct evaluation of $\beta$ by examination of the
transient regime of width growth against time seems to be the least 
accurate procedure, which in this case gave $\beta=0.31(1)$. From scaling, with
$\alpha=0.500(5)$, $z=1.52(3)$, one gets  $\beta=0.33(1)$, in much better agreement
with the exact $\beta=1/3$. Measuring this exponent
directly from the transient regime tends to result in underestimation;
as seen in Secs.~\ref{sec:obcss} and~\ref{sec:qd} above, such a trend
is present in all our subsequent results, both for open systems and for quenched
disorder. Also for PBC, we showed that the dependence of limiting KPZ interface widths
against TASEP particle density can be accounted for by a treatment, which
makes explicit use of the weight factorization that occurs for TASEP with
PBC~\cite{derr93,derr98}. Arguments based on weight factorization provide an 
explanation for the
shape of slope distributions in the interface problem as well.    

In Subsec.~\ref{nr}, we first established that
open-boundary systems in the maximal-current phase, characterized by $\alpha_I$, 
$\beta_E \geq 1/2$, exhibit the same set of KPZ exponents as their PBC counterparts.
For the exponent $z$, this is in agreement with the Bethe ansatz solution~\cite{ess05}.
Examination of interface widths corresponding to systems in the low-density phase, with  
$\alpha_I=1/4$, $\beta_E= 1/2$, shows that curve collapse can be found to a rather 
good extent, giving the following exponent estimates:
$\alpha=0.497(3)$, $z=1.20(5)$, $\beta=0.41(2)$. This apparent
disagreement with the 
Bethe ansatz prediction~\cite{ess05} of $z=0$, which implies a finite 
correlation length, is addressed in Subsec.~\ref{mf} (see two paragraphs on).
We have been able to provide a prediction of the (possibly exact) interface
width exponent $\alpha=1/2$, based on considerations which make explicit use
of a finite correlation length for the TASEP. 

For systems with $\alpha_I=\beta_E <1/2$,
i.e., on the coexistence line of the open-boundary TASEP, interface width scaling gave
$\alpha=0.99(1)$, $z=2.10(5)$, $\beta=0.47(2)$. We provided a prediction
of $\alpha=1$ based on properties of the corresponding TASEP, in this case the fact
that phase separation governs the dominant features of the interface configuration
at stationarity. Direct calculation of density profiles in the particle system
shows the time evolution of a shock (kinematic wave). At late times, the 
ensemble-averaging of local densities tends to mask the evidence of phase segregation,
which can, however, be retrieved by examination of the corresponding interface slope
PDFs.

In Subsec.~\ref{mf}, we outlined a mean-field continuum calculation, which
sheds additional light on the approach to stationarity in open-boundary systems.
We showed that, under suitable conditions such as those at $(\alpha_I,\beta_E)=
(1/4,1/2)$ with uniform initial density $\rho_\ell=1/4$, system-wide propagation
of a kinematic wave  translates into a characteristic time $\propto L^{\,z^\prime}$
($z^\prime=1$ in mean field). This goes towards explaining the apparent inconsistency 
between our result from interface-width evolution, $z=1.20(5)$, and that from the Bethe 
ansatz solution which gives $z=0$. 
Indeed, in that case the $L-$independent relaxation time implied
by $z=0$  is hidden underneath a  slower part of ballistic origin, 
and it is the scaling of the latter which is captured by the interface 
width collapse, but not by considerations solely of the real part of 
Bethe ansatz excitation energies. 
Furthermore, in all other cases where our own
numerical results are consistent with those of Ref.~\onlinecite{ess05},
the mean-field prediction concurs with both. An extension of this work 
has now been carried out using the more precise domain wall method of 
Refs.~\onlinecite{ksks98,ds00} on a kink--like
initial state in the massive phase~\cite{fab-p}. This shows clearly the
ballistic element and the much faster (size-independent) amplitude
decay, providing an independent confirmation of the scenario.

In Sec.~\ref{sec:qd} we investigated quenched bond disorder in TASEP with PBC.
From the scaling analysis of the corresponding interface widths (which in this case
are subjected to correlated, or "columnar" randomness~\cite{k00,ahhk94}) 
we estimate $\alpha=1.05(5)$, $z=1.7(1)$, $\beta=0.62(7)$. For comparison,
values quoted for $\beta$ in standard, two-dimensional quenched disorder in $D=1+1$ 
KPZ systems are close to $0.63$~\cite{csahok93,opz95}. It has been 
argued~\cite{opz95,amaral95} that this type of KPZ model is in the universality class 
of directed percolation, thus one should have $\beta=\beta_{DP}=0.633$.  
Furthermore, in Ref.~\onlinecite{amaral95}, it is shown that by varying the intensity
of the various terms in the quenched counterpart of Eq.~(\ref{eq:kpz}), one can make
KPZ-like systems go through distinct regimes, namely: pinned, with $\alpha_P=0.63(3)$,
$\beta_P=0.67(5)$, $z_P=1.06(8)$; moving,  with $\alpha_M=0.75(4)$, $\beta_M=0.74(6)$, 
$z_M=1.01(10)$; and annealed (i.e., fast-moving interface), with $\alpha_A=0.50(4)$,
$\beta_A=0.30(4)$, $z_A=1.67(26)$. It can be seen that our own set of estimates does
not fully fit into any of these, as could reasonably be expected from the 
extreme correlation between quenched defects which is present here, and not in those 
early examples.

We tested universality properties within the columnar disorder class of models.
This was done by replacing the binary distribution, Eq.~(\ref{eq:distp}), with a 
continuous, uniform one: $P(p)=(1-c)^{-1}$ for $c < p <1$. We used $c=0.1$, 
thereby avoiding the problems associated with allowing
$p=0$, see Ref.~\onlinecite{k00}, while still having a rather
broad distribution. Our data scale similarly to those displayed in
Figure~\ref{fig:qpbc} for Eq.~(\ref{eq:distp}). We get $\alpha=1.05(5)$, 
$z=1.45(10)$
(hence $\beta=0.72(7)$ from scaling), in reasonable agreement with the binary 
disorder case, though error bars for $z$ just fail to overlap. 

We also investigated slope distributions for the quenched disorder problem. These
provide clear evidence of phase separation, a phenomenon known to take place in
such circumstances~\cite{tb98}. 

A direct analytic (mean-field) approach to steady-state
properties of  TASEP with quenched disorder produced closed-form expressions for the 
piecewise shape of averaged profiles densities, as well as rather restrictive bounds on 
currents. All these have been verified in our numerical simulations.
The analytic approach is similar to that of Ref.~\onlinecite{tb98}, 
where it was already shown that a mean field description applies, 
and to part of Ref.~\onlinecite{hs04}. In place of the maximum current
principle used in Ref.~\onlinecite{tb98} we have obtained analytic
consequences of the mean field mapping within segments and combined 
them with segment probabilities to obtain new results, particularly 
for profiles and limiting currents.

We note that,
for weak randomness, characterized by a small value of a disorder parameter $\varepsilon$
[this could be, e.g., $p_s-p_w$ in Eq.~(\ref{eq:distp})], and steady state 
conditions, 
the noiseless (mean-field) constant current condition gives an equation for $\sigma 
\equiv \partial h/\partial x$ in the form:
\begin{equation}
\varepsilon\,\zeta(x)=\frac{\partial\sigma}{\partial x} + \sigma^2\ ,
\label{eq:hmsg}
\end{equation}
where $\zeta(x)$ is a quenched variable corresponding to random bond disorder.
Eq.~(\ref{eq:hmsg}) has the same structure as that found for equations governing
the evolution of local Lyapunov exponents for Heisenberg-Mattis spin glass 
chains~\cite{dqrbs06,dqrbs07}. To see the correspondence, refer, e.g., to Equation~(14)
of Ref.~\onlinecite{dqrbs07}, substituting $\varepsilon$ for (low) magnon frequency 
$\omega$. So one can, in principle, adopt the same type
of Fokker-Planck procedures to find distributions of $\sigma$ in Burgers-like
equations, and hence of slope distributions in KPZ systems.

\begin{acknowledgments}
The authors thank Fabian Essler and Jan de Gier for helpful discussions.
S.L.A.d.Q. thanks the Rudolf Peierls Centre for Theoretical Physics,
Oxford, where most of this work was carried out, for the hospitality,
and CNPq  for funding his visit. The research of S.L.A.d.Q. is financed 
by the Brazilian agencies CNPq (Grant No. 30.6302/2006-3), FAPERJ (Grant
No. E26--100.604/2007), CAPES, and Instituto do Mil\^enio de
Nanotecnologia--CNPq.
R.B.S. acknowledges partial support from EPSRC Oxford Condensed Matter
Theory Programme Grant EP/D050952/1. 
\end{acknowledgments}

\end{document}